\documentclass{emulateapj}

\usepackage{epsfig}

\newcommand{\be}{\begin{equation}}
\newcommand{\ee}{\end{equation}}
\newcommand{\ba}{\begin{eqnarray}}
\newcommand{\ea}{\end{eqnarray}}
\newcommand{\bft}{\mbox{\boldmath $\theta$}}

\begin{document}

\pagestyle{plain}

\title{Size of spectroscopic calibration samples for cosmic shear photometric
redshifts}
\author{Zhaoming Ma{\footnote{Email:mazh@sas.upenn.edu}} and Gary Bernstein}
\affil{Department of Physics and Astronomy,
       University of Pennsylvania, Philadelphia, PA 19104}

\begin{abstract}
Weak gravitational lensing surveys using photometric redshifts can
have their cosmological constraints severely degraded by errors in the
photo-z scale.
We explore the cosmological degradation versus the size of the
spectroscopic survey required to calibrate 
the photo-z probability distribution.
Previous work has assumed a simple Gaussian distribution of photo-z
errors; here, we describe a method for constraining an arbitrary
parametric photo-z error model.  As an example we allow the photo-z
probability distribution to be the sum of $N_g$ Gaussians.  
To limit cosmological degradation to a fixed level, photo-z
distributions comprised of multiple Gaussians
require up to a 5 times larger
calibration sample than one would estimate from assuming a
single-Gaussian model.  This degradation saturates at $N_g\approx 4$
in the simple case where the fiducial distribution is independent of $N_g$.
Assuming a single Gaussian when the photo-z distribution has multiple
parameters underestimates cosmological parameter uncertainties by up to $35\%$.
The size of required calibration sample also depends on the shape of the
fiducial distribution, even when the rms photo-z error is held fixed.
The required calibration sample size varies up to a factor of 40 among
the fiducial models studied, but this is reduced to a factor of a few if the
photo-z error distributions are forced to be slowly varying with redshift.
Finally, we show that the size of the required calibration sample can
be substantially reduced by optimizing its redshift distribution.
We hope this study will help stimulate work on better understanding of
photo-z errors.
\end{abstract}

\section{Introduction}
\label{sec:introduction}

Explaining the Hubble acceleration, {\it i.e.} the ``dark energy,''
is one of the main challenges to cosmologists. Weak gravitational
lensing (WL) has perhaps the most potential to constrain dark energy
parameters of any observational window, but is a newly developed
technique which could be badly degraded by systematic errors 
(Albrecht et al 2005).  A WL survey requires an estimate of the shape 
and the redshift
of each source; dominant observational systematic errors are expected
to be errors in galaxy shape due to the uncorrected influence of the point
spread function (PSF) and errors in estimation of redshift
distributions if they are determined by photometric redshifts (photo-z's).
Interpretation of WL data could also be systematically incorrect due
to errors in the theory of the non-linear matter power spectrum or
intrinsic alignments of galaxies.  In this paper we present a new and
more general analysis of the effect of photo-z calibration errors and
of the size of the spectroscopic survey required to reduce photo-z errors
to a desired level.

Recent work has addressed many of these potential systematic errors in
WL data and theory:
from the computation of the nonlinear matter power spectrum 
\citep{Vale_White, White_Vale, LosAlamos, Huterer_Takada, Hagan_Ma_Kravtsov,
Linder05, Ma06, Francis07};
from baryonic cooling and pressure forces on the distribution of large-scale 
structures \citep{White_baryons, Zhan_Knox, Jing06, Rudd07, Zentner07};
approximations in inferring the shear from
the maps \citep{Dod_Zhang, White_reduced, Dod_Shapiro, Shapiro06}; 
and the presence of dust \citep{Vale_Hoekstra}.
The promise and problems of WL have stimulated work on how to improve 
the PSF reconstruction \citep{Jarvis_Jain}, estimate shear from noisy images
\citep{Bernstein_Jarvis,Hirata_Seljak,Hoekstra04,Heymans06,
Nakajima06,STEP2_07}, 
and protect against errors in the theoretical power spectrum at small scales
\citep{nulling}.

For visible-NIR WL galaxy surveys, the dominant systematic error is
likely to be inaccuracies in the photo-z calibration.
The effect of photo-z calibration on weak lensing is studied by
\cite{Ma05, Huterer05_wlsys, Jain06, Abdalla07}; and \cite{Bridle07}.
The distributions of photo-z errors assumed for these studies are,
however, much simpler than will exist in real surveys
\citep{Dahlen07,Oyaizu07,Wittman07,Stabenau07}.
\citet{Huterer05_wlsys} assumed that photo-z errors take the form of
simple shifts (a bias that varies with $z$), while \cite{Ma05} assume
the photo-z error 
distribution is a Gaussian, with a bias {\em and} dispersion that are
functions of $z$.  These studies
find that dark energy
constraints are very sensitive to the uncertainties of photo-z parameters.
A spectroscopic calibration sample of galaxies on the order of $10^5$ is
required to have less than $50\%$ degradation on dark energy constraints.
In this work we relax the Gaussian assumption, presenting a method to
evaluate the degradation of dark energy parameter accuracy versus the size
of the spectroscopic calibration survey, for the case of a photo-z error
distribution described by any parameterized function.  We then apply
this to a model in which the core of the photo-z error distribution is
the sum of multiple Gaussians, ignoring for now the effect of
so-called catastrophic photo-z errors or outliers.

    The outline of the paper is as follows. In {\S}\ref{sec:methodology}
we introduce the formalism and parameterizations of cosmology, galaxy
redshift distributions and photometric redshift errors. The implementation
of the formalism is detailed in {\S}\ref{sec:implementation}.
We show the dependence of the size of the calibration sample on the
number of Gaussians and the shapes of the fiducial photo-z models in 
{\S}\ref{sec:size}. We illustrate the effectiveness of optimizing the
calibration sample in {\S}\ref{sec:optimize}. We discuss our results and
conclude in {\S}\ref{sec:conclusion}.

\section{Methodology}
\label{sec:methodology}

    Two major generalizations are made to the work done in \cite{Ma05}.
One is that we do {\em not} assume {\it a priori} knowledge of the 
true underlying (unobserved) galaxy redshift distribution $n(z)$. Instead,
we treat it as an unknown function which must be constrained by the
photo-z distribution  $n(z_{\rm ph})$ and other observables.
The other modification we make is to generalize the 
photo-z probability distribution to generic parametric functions, in
our case multiple Gaussians.

\subsection{Galaxy Redshift Distributions and Parameters}

   One of the observables that a weak lensing survey would provide is the
galaxy photo-z distribution $n(z_{\rm ph})$. The corresponding 
galaxy true redshift distribution $n(z)$ is unknown. These two 
galaxy redshift distributions are related by the photo-z probability 
distribution $P(z_{\rm ph}|z)$,
\begin{equation}
  n(z_{\rm ph})=\int n(z)P(z_{\rm ph}|z)dz \,.
\label{eq:nzpnz}
\end{equation}
In practice, we model the true $n(z)$ as a linear interpolation
between values $n^i$ at a discrete set of redshifts $\{z^i\}$.  The
$n^i$ become free parameters in a fit to the observables.

Weak-lensing tomography \citep{Hu99, Huterer02} extracts temporal 
information by dividing $n(z_{\rm ph})$ into a few
photo-z bins. The true distribution of galaxies $n_i(z)$ that fall in the
$i$th photo-z bin with $z_{\rm ph}^{(i)} < z_{\rm ph} < z_{\rm ph}^{(i+1)}$
becomes
\begin{equation}
   n_i(z) = \int_{z_{\rm{ph}}^{(i)}}^{z_{\rm{ph}}^{(i+1)}}
            dz_{\rm{ph}} \, n(z)\, P(z_{\rm{ph}} | z)\,.
\label{eq:ni}
\end{equation}
\cite{Ma05} had taken $P(z_{\rm{ph}} | z)$ to be a Gaussian,
described by two parameters (redshift bias and rms) at a given value of $z$.
Now we allow a generic dependence on a set of photo-z parameters $p_\mu$
indexed by $\mu$, $P(z_{\rm ph} | z; p_\mu)$. For a multiple Gaussian photo-z
model, $p_\mu$ are the biases and rms values of the component Gaussians.

The  total number of galaxies per
steradian
\be
  n^A = \int_0^\infty dz n(z) \,,
\ee
fixes the normalization, and we analogously define
\be
  n^A_{i} = \int_0^\infty dz n_{i}(z) \,
\ee
for the tomographic bins.

\subsection{Observables}

    We utilize information from both lensing and redshift surveys which
include galaxy photo-z distribution and the spectroscopic calibration
sample for the photo-z's.

\subsubsection{Lensing Cross Spectra}
   Following \cite{Ma05}, we choose the number-weighted convergence
power spectra $n^A_i n^A_j P_{ij}^{\kappa}(\ell)$ as lensing 
observables{\footnote{Since we are using all the information from the galaxy number
distribution in this study, one could equally well use $P_{ij}$ as
lensing observables.}},  
where $i$ and $j$ label tomographic bins. From \cite{Kaiser_92, Kaiser_98}
we have
\begin{equation}
n^A_i n^A_j P_{ij}^{\kappa}(\ell) =
\int_0^{\infty} dz \,{W_i(z)\,W_j(z)}{ H(z) \over D^2(z)}\,
 P\! \left (k_{\ell},  z\right ),
\label{eq:pk_l}
\end{equation}
\noindent where $H(z)$ is the Hubble parameter, $D(z)$ is the angular
diameter distance in comoving coordinates, $P(k_{\ell}, z)$ is the
three-dimensional matter power spectrum, and $k_{\ell} = \ell /D(z)$ is the
wavenumber that projects onto the multipole $\ell$ at redshift $z$.  The
weights $W$ are given by
\begin{eqnarray}
 W_i(z) &=& {3\over 2}\,\Omega_m\, {H_0^2 D(z) \over H(z)}(1+z)\nonumber  \\
 &&\times
 \int_z^\infty
 dz^{\prime} {n_i(z^{\prime})}
{D_{LS}(z,z') \over D(z')} \,,
\label{eqn:weights}
 \end{eqnarray}
 where $D_{LS}(z,z')$ is the angular diameter distance between the two
redshifts.  We compute a power spectrum from the transfer function of
\cite{Hu_transfer} with dark energy modifications from \cite{Hu01c}, and the
nonlinear fitting function of \cite{PD96}.

\subsubsection{Photo-z Distribution}
   Another set of observables from the redshift surveys is the galaxy photo-z
distribution, $n(z_{\rm ph}^i)$, collected into
bins.  The width $\delta z_{\rm ph}$ of these bins would typically be much
finer than the tomography 
bins and should be at least as fine as the nodes $z^i$ on
which the true redshift distribution is defined.  
Binning equation\,\ref{eq:nzpnz}, we have
\begin{equation}
  n(z_{\rm ph}^i) \delta z_{\rm ph} =\int n(z)P(z_{\rm ph}|z; p_\mu) \delta z_{\rm ph} dz \,.
\label{eq:nzpbin}
\end{equation}
So the observables are functions of the intrinsic distribution
$\{n^i\}$ and the photo-z parameters $p_\mu$.

\subsubsection{Spectroscopic Redshifts}
The last piece of information we utilize is the spectroscopic calibration
sample. We presume that a representative sample of $N_{\rm spect}^i$
galaxies has been drawn from the sources in redshift bin $i$, with
spectroscopic redshifts 
determined for {\em all} of them.  Equivalently, we can demand that the
failure rate for obtaining redshifts in the spectroscopic survey must
be completely independent of redshift.
The likelihood of the $j$th
spectroscopic survey galaxy with photo-z value $z_{\rm ph}^j$ being
observed to have spectroscopic redshift $z^j$ is of course
$P(z_{\rm ph}^j|z^j; p_\mu).$  Each spectroscopic redshift hence
adds a little more constraint to the photo-z parameters, as quantified
in \S\ref{sec:speczfish} below.  We presume
all the spectroscopic $z$ values are independent, {\it i.e.} we ignore
source clustering.  While this may be unrealistic in practice for
spectroscopic surveys over small areas of sky, it is
more likely---and adequate---that the redshift {\em errors} are
uncorrelated, so that we can constrain
$P(z_{\rm ph}-z | z)$ with $N^i_{\rm spect}$ independent samples.

We have considered the spectroscopic sample to constrain $P(z_{\rm
  ph}|z)$, which can combine with photo-z counts $n(z^i_{\rm ph})$ to
constrain the true redshift distribution $n(z)$.
One could potentially assume the spectroscopic sample to sample and
constrain $n(z)$ directly.
We avoid this for two reasons. First, claiming both uses for the
spectroscopic sample would be ``double counting'' its information.
Second, a direct constraint of $n(z)$ 
would depend heavily on the assumption that the
calibration sample is a fair representation of the full photo-z sample.
Source clustering in the spectroscopic sample would be more of an
issue.  In addition, we investigate below the possibility of 
targeting calibration samples at rates that vary with redshift.
In this situation, the calibration sample could deviate
from the true underlying galaxy redshift distribution by quite a bit.

It remains crucial, in any case, that
the calibration sample is a fair representation of the 
photo-z sample {\em within each redshift bin and for every galaxy type.}
For example, if we are taking spectra for $5\%$ of the photo-z sample
in some redshift bin, we must be sure to 
draw $5\%$ of the blue galaxies and $5\%$ of the red galaxies for our
complete spectroscopic survey and succeed in obtaining redshifts for
all regardless of color.

\subsection{Fisher Matrix}

    The Fisher matrix quantifies the information contained in the 
observables. The total Fisher matrix is the sum of that from each of
three kinds of (uncorrelated) observables: the lensing shear, the
observed photo-z distribution, and the spectroscopic redshift distribution,
\be
   F_{\mu \nu}^{total} = F_{\mu \nu}^{lens} + F_{\mu \nu}^{n(z_{\rm ph})}
                        +F_{\mu \nu}^{spect} \,,
\label{eqn:fisherTotal}
\ee
and the errors on the parameters are given by 
$\Delta p_{\mu} = [{\bf F}^{total}]^{-1/2}_{\mu\mu}$.

\subsubsection{Lensing Cross Spectra}
    The ${\bf F}^{lens}$ quantifies the information contained in the
lensing observables
\be
O_a(\ell) = n^A_i n^A_j P_{ij}^{\kappa}(\ell)\,, \quad (a\equiv\{ij\},
\; i\ge j)
\ee
on a set of cosmological, photo-$z$ parameters $p_{\mu}$ and the underlying
galaxy redshift distribution parameters.  Under the
approximation that the shear fields are Gaussian out to $\ell_{\rm max}$,
the Fisher matrix is given by
\be
F_{\mu \nu}^{lens} = \sum_{\ell=2}^{\ell_{\rm max}}(2\ell+1)f_{\rm sky}{\sum_{ab}
             {\partial{O_{a}}
            \over \partial{p_{\mu}} } [ {\bf C}^{-1} ]_{ab}
             {\partial{O_{b}}
            \over \partial{p_{\nu}}  }}\,.
\label{eq:fisherLens}
\ee

    Given shot and Gaussian sample variance, the covariance matrix of the
observables becomes
\be
C_{ab} =
n^{A}_{i} n^{A}_{j} n^{A}_{k} n^{A}_{l} \left(
P^{\rm tot}_{ik}  P^{\rm tot}_{jl} +
        P^{\rm tot}_{il}  P^{\rm tot}_{jk}\right)\,,
\label{eq:Cov}
\ee
where $a\equiv\{ij\}$ and $b\equiv\{kl\}$. The total power spectrum is given by
\begin{equation}
P^{\rm tot}_{ij}=P_{ij}^{\kappa} +
\delta_{ij} {  \gamma_{\rm int}^2  \over {n}_i^A} \,,
\label{eq:C_obs}
\end{equation}
where $\gamma_{\rm int}$ is the rms shear error per galaxy per component
contributed by intrinsic ellipticity and measurement error.
For illustrative purposes we use $\ell_{\rm max}=3000$, $f_{\rm sky}$
corresponding to $20,000\,{\rm deg}^2$, $\bar n^{A}$ corresponding to
30 galaxies arcmin$^{-2}$, and $\gamma_{\rm int}=0.22$.  This is what
might be expected from an ambitious ground-based survey like the
Large Synoptic Survey Telescope (LSST).\footnote{{See http://www.lssto.org}}

    For the cosmological parameters, we consider four parameters that 
affect the matter power spectrum: the physical matter density 
$\Omega_{m} h^{2} (= 0.14)$, physical baryon density
$\Omega_{b} h^{2} (= 0.024)$, tilt $n_{s} (= 1)$, and the amplitude
$\delta_{\zeta}(=5.07 \times 10^{-5}$ ; or $A= 0.933$ \citet{Speetal03}).
Values in parentheses are those of the fiducial model. 
To these four cosmological parameters, we add three dark
energy parameters: the dark energy density $\Omega_{\rm DE}(=0.73)$,
its equation of state today $w_{0} = p_{\rm DE}/\rho_{\rm DE}|_{z=0} (=-1)$
and its derivative $w_{a} = -dw/da |_{z=0} (=0)$
assuming a linear evolution with the scale factor $w = w_{0} + (1-a)w_{a}$.
Unless otherwise stated, we shall take {\it Planck} priors on these
seven parameters (W. Hu, private communication).

\subsubsection{Photo-z Distribution}
    The ${\bf F}^{n(z_{\rm ph})}$ quantifies the information contained
in the galaxy photo-z distribution.  We use the model of
equation~(\ref{eq:nzpbin}) to find the dependence of each observable
$n(z_{\rm ph}^i)$ on the true redshift and photo-z parameters.  Each
bin is presumed to have Poisson uncertainties
\be
   \sigma(n(z_{\rm ph}^i) \delta z_{\rm ph}) = 
         [n(z_{\rm ph}^i) \delta z_{\rm ph}]^{\onehalf}.
\ee
In practice, the number of photo-z's will be large, and ${\bf F}^{n(z_{\rm
ph})}$ acts like a linear constraint on the other parameters.

\subsubsection{Spectroscopic Redshifts}
\label{sec:speczfish}
    The ${\bf F}^{spect}$ quantifies the information contained in the
spectroscopic calibration sample on photo-z parameters $p_{\mu}$.  The
simple likelihood analysis of Appendix~A shows that the Fisher matrix
from the spectroscopic survey is
\begin{equation}
  F_{\mu \nu}^{spect} = \sum_i N_{\rm spect}^{i} \int {dz_{\rm ph} {1
      \over {P^i(z_{\rm ph}|z)}}
                {\partial P^i \over {\partial p_{\mu}}}
                {\partial P^i \over {\partial p_{\nu}}} } \,,
\label{eqn:NspecPrior}
\end{equation}
where $N^i_{\rm spect}$ spectra have been obtained from redshift bin
$i$ (out of $N_{\rm pz}$) and $P^i$ describes the photo-z errors for
this bin.

\section{Implementation}
\label{sec:implementation}
We now apply the above formalism to derive Fisher matrices for
specific cases of WL surveys and their associated spectroscopic
calibration surveys.  In further sections we vary the parameters of
the photo-z errors and the spectroscopic survey and investigate the
impact on the accuracy of dark energy parameters derived from each survey.

Following \cite{Ma05}, the fiducial galaxy redshift distribution $n(z)$
is chosen to have the form
\begin{equation}
   n(z) \varpropto {z}^{\alpha} \exp\left [-(z/z_0)^\beta\right ] \,.
\label{eq:nz}
\end{equation}
Unless otherwise stated we adopt $\alpha=2$ and $\beta=1$ and fix $z_0$ such
that the median redshift is $z_{\rm med} = 1$.  The parametric model for
$n(z)$ is determined by linear interpolation between $N_{\rm pz}=31$ values
$n^i=n(z^i)$ at equally spaced redshifts between 0 and 3.

In the Gaussian case as assumed in \cite{Ma05}, we have
\begin{equation}
  P(z_{\rm ph}|z) = {1 \over {\sqrt{2 \pi} \sigma_z}} \exp \left [ 
     {-{{(z_{\rm ph} - z - z_{\rm {bias}})^2} \over {2 \sigma_z^2}}} \right ] \,.
\end{equation}
The bias $z_{\rm bias}$ and dispersion $\sigma_z$ are functions of
$z$.

In reality, $P(z_{\rm ph}|z)$ could be far more complex than a single Gaussian.
We explore this complexity by assuming $P(z_{\rm ph}|z)$ as the sum of
Gaussians. Using $N_{g}$ Gaussians to describe $P(z_{\rm ph}|z)$, we have
\begin{eqnarray}
  P(z_{\rm ph}|z) &=&{} \sum_{j=1}^{N_{g}}
            {C_j \over {\sqrt{2 \pi} \sigma_{z;j}}} 
     \nonumber \\
  &&{}  \times \exp \left [ 
     {-{{(z_{\rm ph} - z - z_{\rm {bias;j}})^2} \over {2 \sigma_{z;j}^2}}} 
          \right ]  \,,
\label{eq:PnG}
\end{eqnarray}
where $C_j$ is the normalization of the $j^{th}$ Gaussian. Since we assume
$P(z_{\rm ph}|z)$ is normalized to unity, we have $\sum_j C_j = 1$. We allow
the biases $z_{\rm {bias;j}}(z)$ and scatters $\sigma_{z;j}(z)$ to be arbitrary 
functions of redshift. The redshift distribution of the tomographic bins
defined by equation\,\ref{eq:ni} can then be written as
\begin{eqnarray}
 n_i(z) &=&{} {1 \over 2} n(z)\, \sum_j^{N_g} C_j[ {\rm erf}(x_{i+1;j}) - 
                 {\rm erf}(x_{i;j}) ],
 \end{eqnarray}
 with
\begin{eqnarray}
   x_{i,j} &\equiv & ({z_{\rm{ph}}^{(i)}-z+z_{\rm{bias;j}}})/
                {\sqrt{2} \sigma_{{\rm z};j}},
\end{eqnarray}
where ${\rm erf}(x)$ is the error function.

In practice, we represent the free functions $z_{\rm{bias;j}}(z)$ and
$\sigma_{z;j}(z)$ by linear interpolation between values at a discrete
set of $N_{\rm pz}$ redshifts equally 
spaced from $z=0$ to 3.  The photo-z parameter set $\{p_\mu\}$
is hence the $2N_gN_{\rm pz}$ values of the biases and dispersions of
the Gaussians at these nodes.

\begin{figure*}[ht]
\centerline{\psfig{file=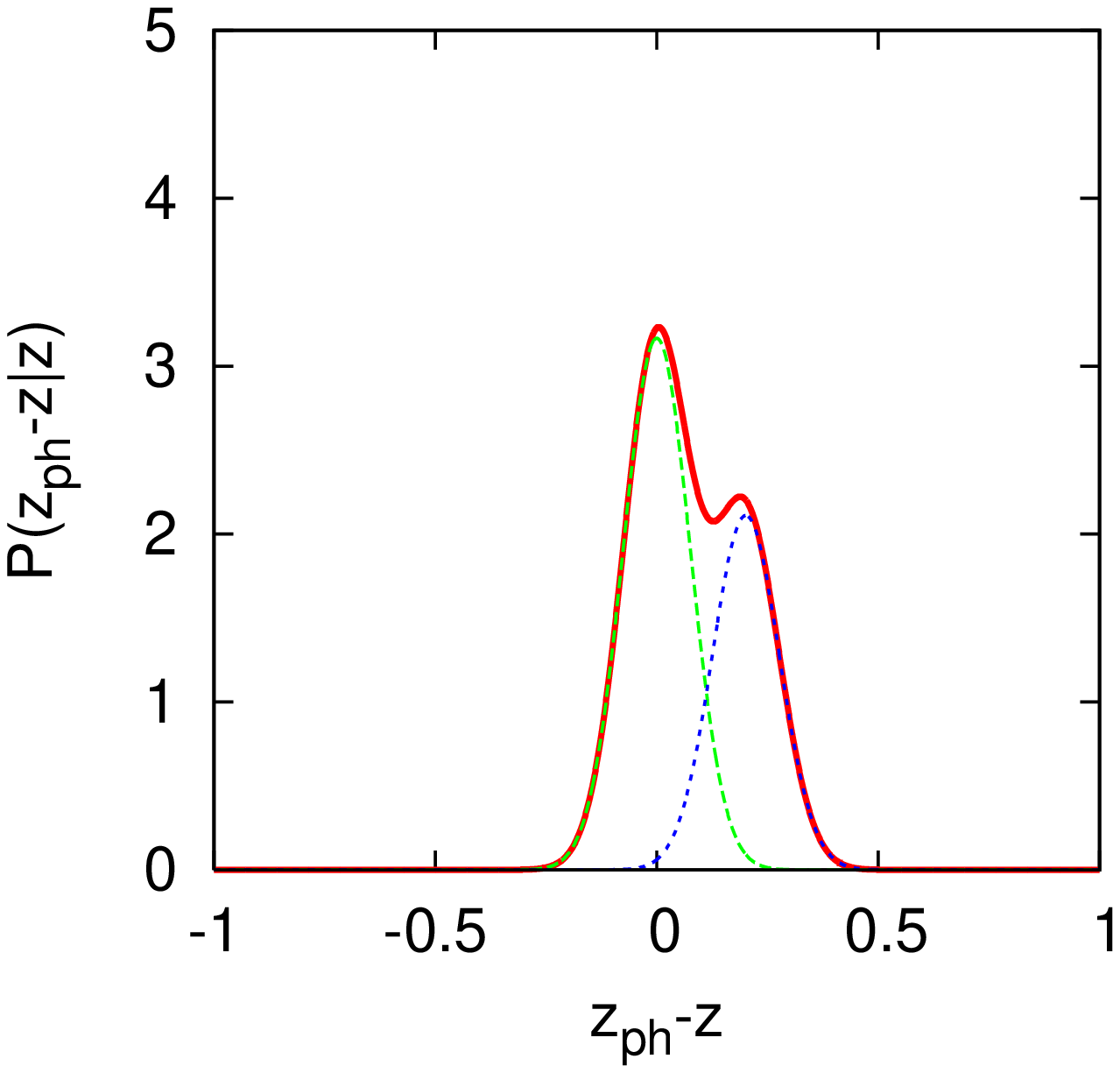, width=1.6in}
            \psfig{file=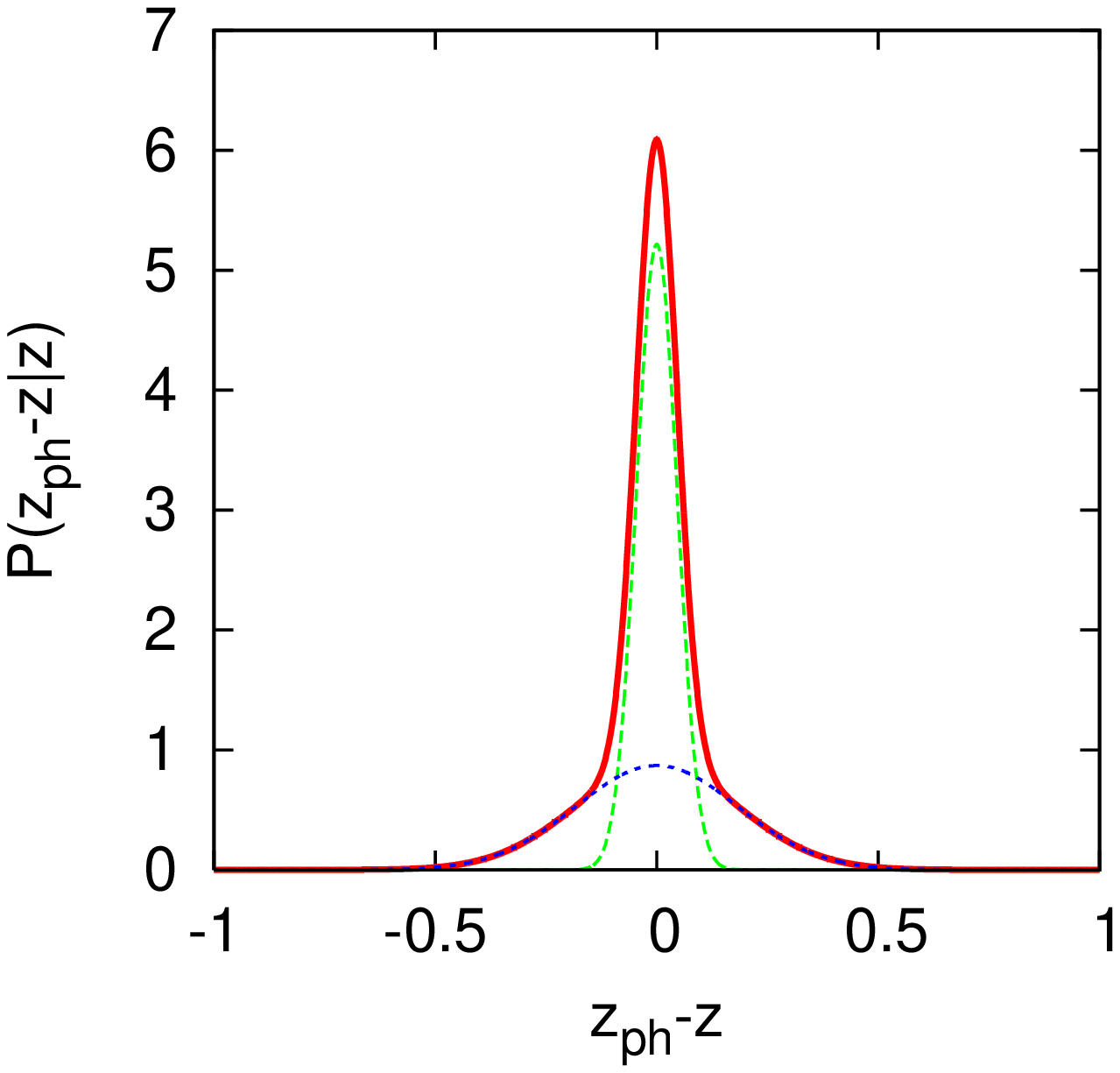,  width=1.6in}
            \psfig{file=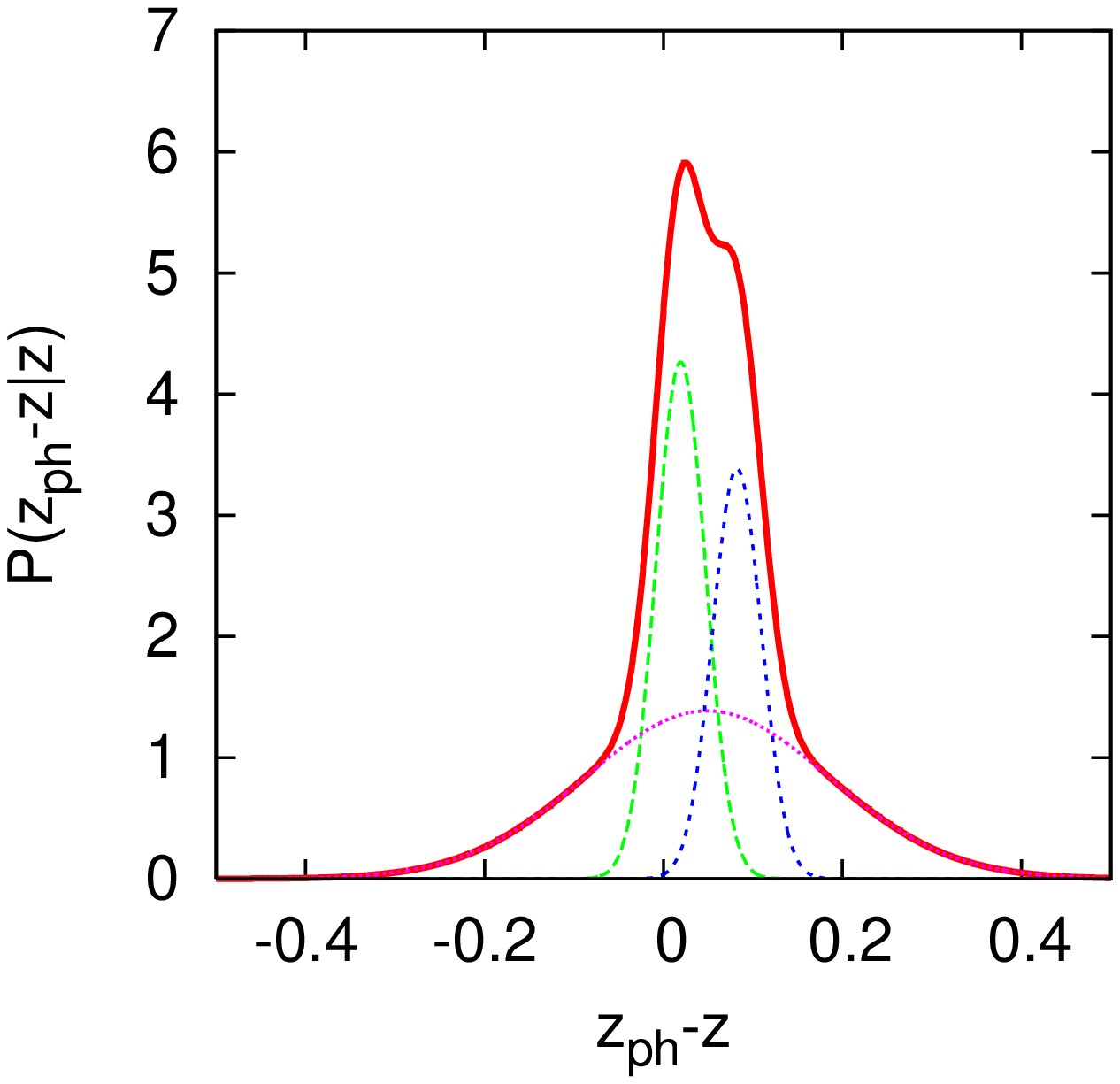,width=1.6in}
            \psfig{file=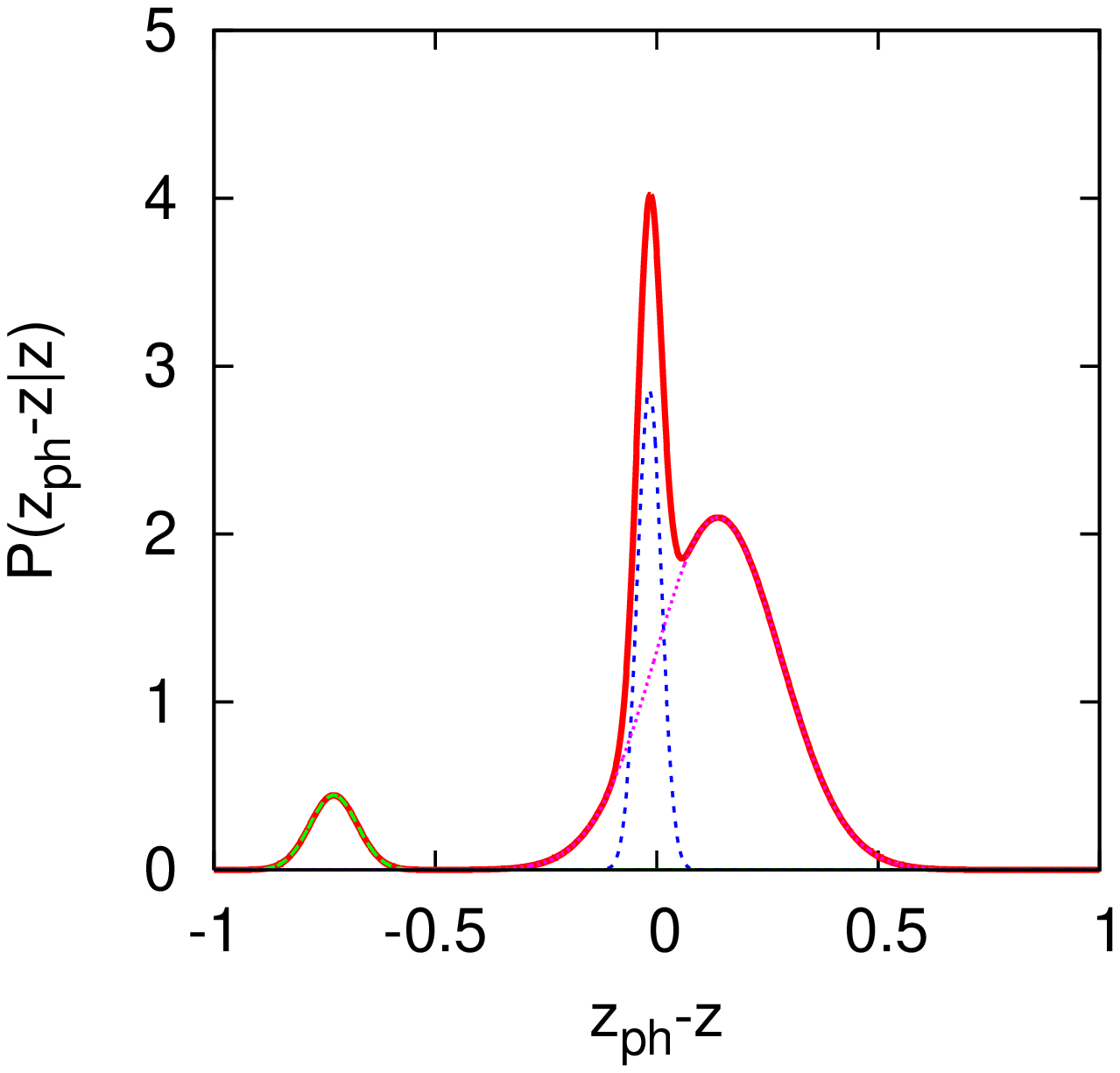,  width=1.6in} }
\caption {Examples of photo-z probability distribution $P(z_{\rm ph}|z)$.
          From left to right, they are two Gaussians with different biases,
          two Gaussians with different $\sigma_z$ values, three Gaussians with
          parameters randomly generated, and three Gaussians with one being
          catastrophic. The thick solid lines are the total $P(z_{\rm ph}|z)$,
          and the thin dotted lines are the individual Gaussians that build
          up $P(z_{\rm ph}|z)$.
          }
\label{fig:pzPDFeg}
\end{figure*}

   With multiple Gaussians, we can describe a wide variety of  photo-z 
probability distributions $P(z_{\rm ph}|z)$. Figure\,\ref{fig:pzPDFeg}
shows a few examples of $P(z_{\rm ph}|z)$.  A wide variety of
behaviors can be represented, including ``catastrophic'' outliers.
Although catastrophic photo-z errors could potentially have a big impact
on what we can get out of cosmic shear surveys \citep{Amara06},
we restrict ourselves to studying the core of
$P(z_{\rm ph}|z)$ in this study.

\cite{Ma05} show that $N_{\rm pz}=31$ between $z=0$ and 3 gives
enough freedom to the photo-z parameters to destroy all tomographic information. 
Since we are giving the photo-z even more freedom by allowing
$P(z_{\rm ph}|z)$ to be  
multiple Gaussians, $N_{\rm pz}=31$ should be large enough. Unless stated 
otherwise, we use $N_{\rm pz}=31$. Thus, the total number of photo-z 
parameters is $62 N_g$.

The observables 
$n(z_{\rm ph}^i)$, determined in bins of width
$\delta z_{\rm ph}$, need not have the same bin width as the spacing
of the $n(z^i)$ or the photo-z parameters.  In fact, they should be
more finely spaced.
We choose the size of $\delta z_{\rm ph}$ 
such that further dividing it by two does not lead to anymore information 
gains. We find that $\delta z_{\rm ph} = 0.0125$ is small enough for all 
the photo-z models explored in this study.

\section{Size of the Spectroscopic Calibration Sample}
\label{sec:size}

In this section we investigate the size of
the spectroscopic calibration sample required to limit photo-z
systematics to some desired level. In particular, we are interested in
the increased demands that might result from giving the photo-z
distribution freedom to depart from a single-Gaussian form.
We first demonstrate that, for a fixed
fiducial photo-z model, the required calibration size increases with
the number of degrees of freedom ($2N_g$) that we allow for deviations
from the fiducial model. This increase reaches an asymptotic limit
with $N_g$.

Second, we investigate how the required $N_{\rm spect}$ varies as
we allow the fiducial model to assume non-Gaussian shapes.
Equations\,\ref{eqn:NpriorGmu} and \ref{eqn:NpriorGsi} show that in 
the case of a Gaussian distribution, the $N_{\rm spect}$ required to
constrain the photo-z parameters is proportional to 
the square of the width of the distribution. In the following, we hold 
the width (defined as the rms)
of the fiducial photo-z distributions to be $0.05(1+z)$. Holding this
fiducial width fixed means that any variations we
see are due only to variations in the {\em shape}
of the photo-z probability distribution.

We use the error degradations in $w_{\rm a}$ (that is, errors in 
$w_{\rm a}$ relative to the error with perfect knowledge of the photo-z
parameters) as the measure of dark energy degradations. The error
degradations in $w_{\rm p}$\footnote{We have $w_{\rm p} \equiv w(z=z_{\rm p})$,
where $z_{\rm p}$ is the redshift at which the errors of $w_0$
and $w_{\rm a}$ are decorrelated.}
are about $30$-$50\%$ lower and follow the same trend as that of $w_{\rm a}$.
Roughly speaking, the figure of merit adopted by the Dark Energy Task
Force \citep{DETF} will degrade as the square of the dark energy
degradation used here.

In this section we assume that the $N_{\rm spect}$ total
spectroscopic galaxies are selected uniformly in redshift
between 0 and 3.

\subsection{Dependence on the Number of Gaussians $N_g$}
\begin{figure*}[ht]
\centerline{\psfig{file=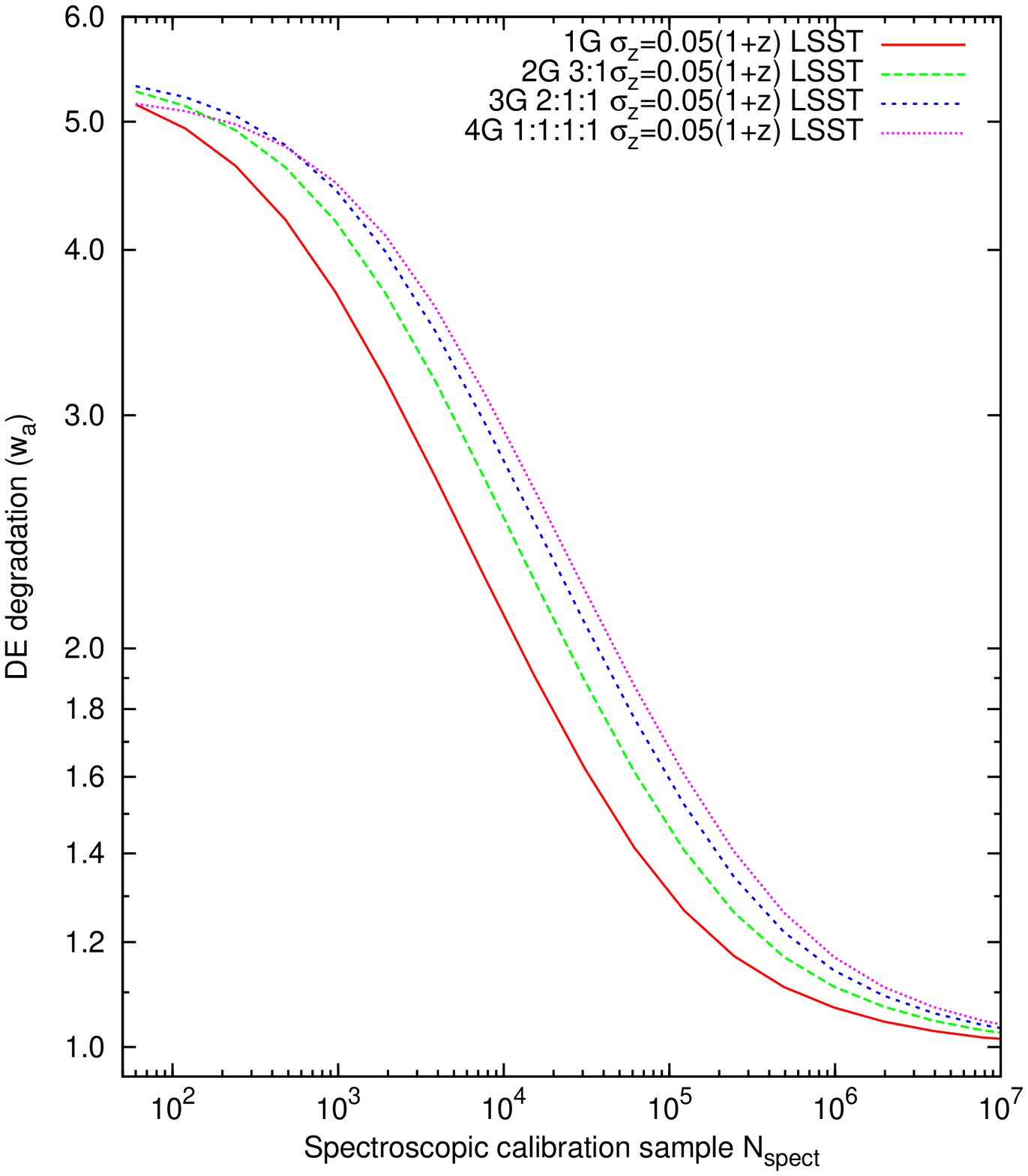, width=3.2in}
            \psfig{file=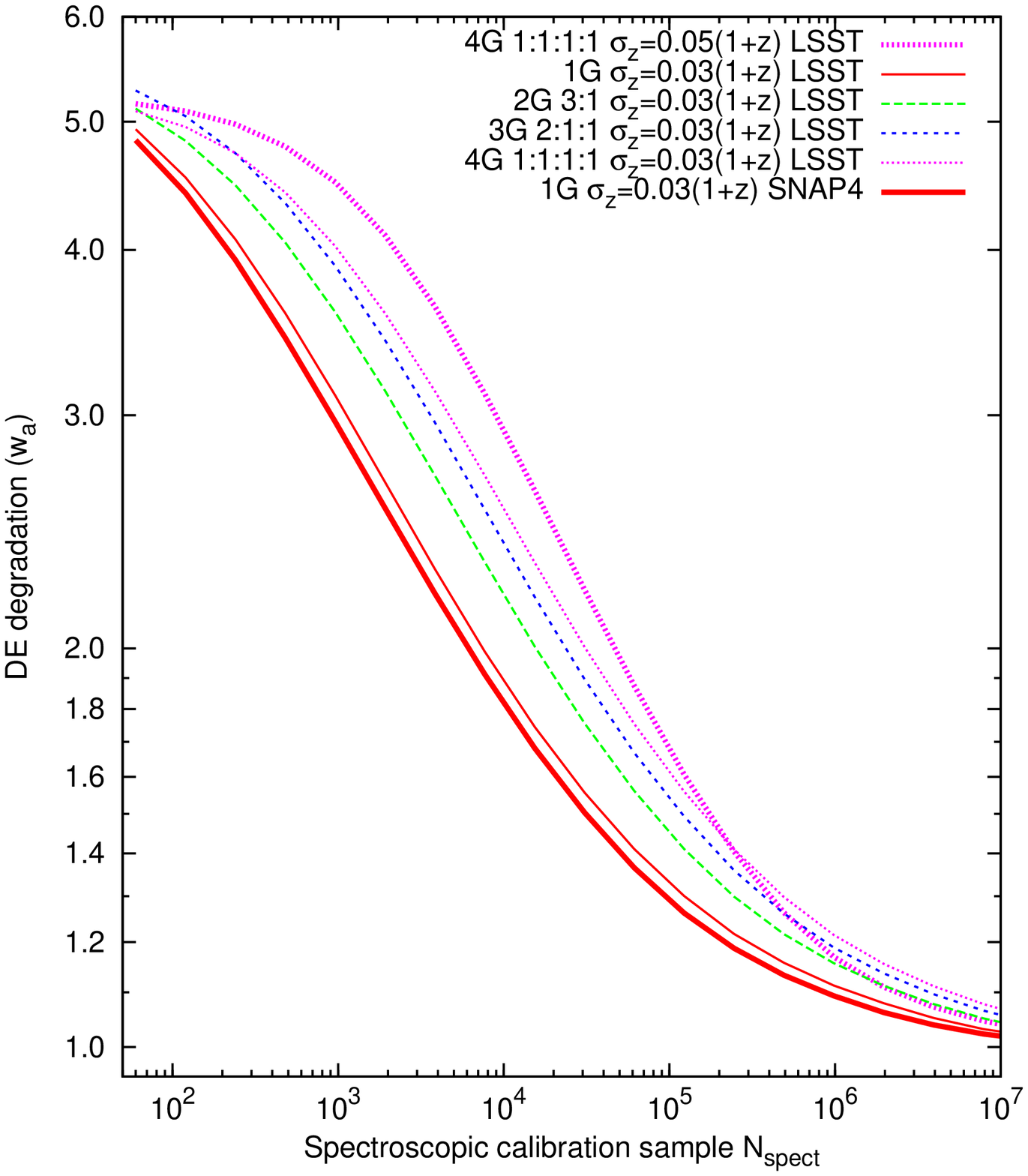, width=3.2in}   }
\caption {Left:
          The $N_{\rm spect}$ requirement for the same fiducial photo-z
          distribution being
          modeled using different numbers of Gaussians. These Gaussians
          differ only by their normalizations, whose ratios are shown in
          the legend (e.g., ``2G 3:1'' means a
          two-Gaussian model with normalization ratio of 3 to 1.).
          All Gaussian components have fiducial $z_{\rm bias} = 0$ and 
          $\sigma_z = 0.05(1+z)$, so all cases have the same fiducial
          distribution while the fitted photo-z error distribution
          gains more freedom.
          The solid line is for the case of single
          Gaussian ($N_g = 1$). 
          Survey specs are LSST-like.
          Right:
          Thin lines are the same as those in the left panel but
          with $\sigma_z = 0.03(1+z)$.
          For comparison, the four-Gaussian model in the left panel
          is plotted as the thick dotted line (magenta). The thick
          solid line (red) is the single-Gaussian model with {\it SNAP}-like
          survey specs ($f_{\rm sky} = 4000$ deg$^2$, $\bar n^{A}=100$ 
          galaxies arcmin$^{-2}$, and $\gamma_{\rm int}=0.22$).
          }
\label{fig:fid1234G}
\end{figure*}

The left panel of Figure\,\ref{fig:fid1234G} plots the 
dark energy degradation versus the 
size of the spectroscopic calibration sample, when the
photo-z error distribution has
$N_g = 1$, $2$, $3$, and $4$.  The fiducial biases and dispersions are
the same for all component Gaussians.  So the fiducial $P(z_{\rm
  ph}|z)$ is identical in all cases, but with higher $N_g$, there is
more freedom for deviations from the fiducial.  The second, third, and
fourth Gaussian components are each fixed to have one-fourth the total
normalization of the distribution.

At fixed dark energy degradation, the 
required size of the calibration sample ($N_{\rm spect}$) increases with 
the number of Gaussians and reaches an asymptotic value when $N_g \approx 4$. 
When dark energy degradation is 1.5, the $N_{\rm g} = 4$ photo-z model requires
$\approx5$ times the calibration sample of the $N_{\rm g} = 1$ model.

Another view is that the dark-energy uncertainties will be
underestimated if one fits a single-Gaussian model to photo-z
distributions that actually require more freedom.
For example, assume we obtain
$4 \times 10^4$ spectra, as required to
keep dark energy degradation under 1.5 for a single-Gaussian photo-z
model. We find, however that the dark energy
degradation for $N_g=4$ rises above 2.0.  So relaxing the Gaussian
assumption for photo-z's inflates the cosmological uncertainties by
$\approx35\%$. 

We also note from the left panel of Figure~\ref{fig:fid1234G} that
the dark energy
degradation has a characteristic dependence on $N_{\rm spect}$; for
$N_{\rm spect}\gtrsim10^3$, the dark energy parameter error scales roughly as
$N_{\rm spect}^{1/4}$.  When the dark energy degradation reaches $\approx1.2$,
at $N_{\rm spect}=10^5$--$10^6$, the gains from additional spectra
become weaker and a degradation of unity is approached only very
slowly.  As we vary $N_g$, we change the location of this ``knee'' in
the curve, but not the scaling for $N_{\rm spect}$ below the knee.
This scaling is not sensitive to either the fiducial photo-z models
or survey specs. For example, as shown in the right panel of
 Figure~\ref{fig:fid1234G}, for a photo-z model with 
$\sigma_z = 0.03 (1+z)$, the scaling is $N_{\rm spect}^{1/5}$;
for a SNAP-like survey{\footnote{See http://snap.lbl.gov}} with 
$f_{\rm sky} = 4000$ deg$^2$, $\bar n^{A}=100$
galaxies arcmin$^{-2}$, and $\gamma_{\rm int}=0.22$,
the scaling is also $N_{\rm spect}^{1/5}$
as shown in the right panel of Figure~\ref{fig:fid1234G}.


The desired spectroscopic survey size $N_{\rm spect}$ will in general
depend on the width and shape of the fiducial photo-z distribution,
not just $N_g$.  We next investigate the dependence on the detailed
shape of the fiducial distribution.

\subsection{Dependence on the Fiducial Photo-z Models}
\label{sec4.2}

\begin{figure*}[ht]
\centerline{\psfig{file=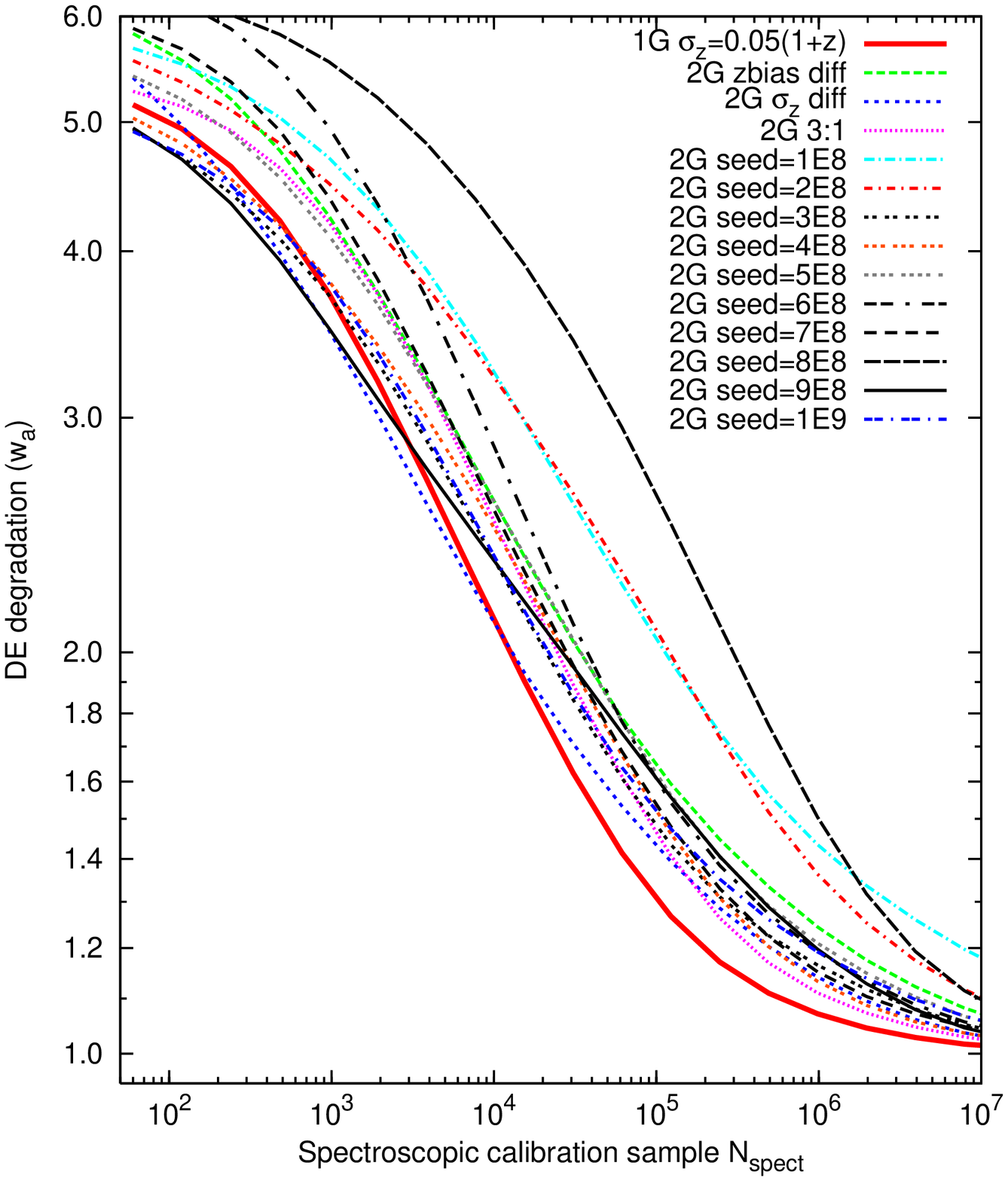, width=2.1in}
            \psfig{file=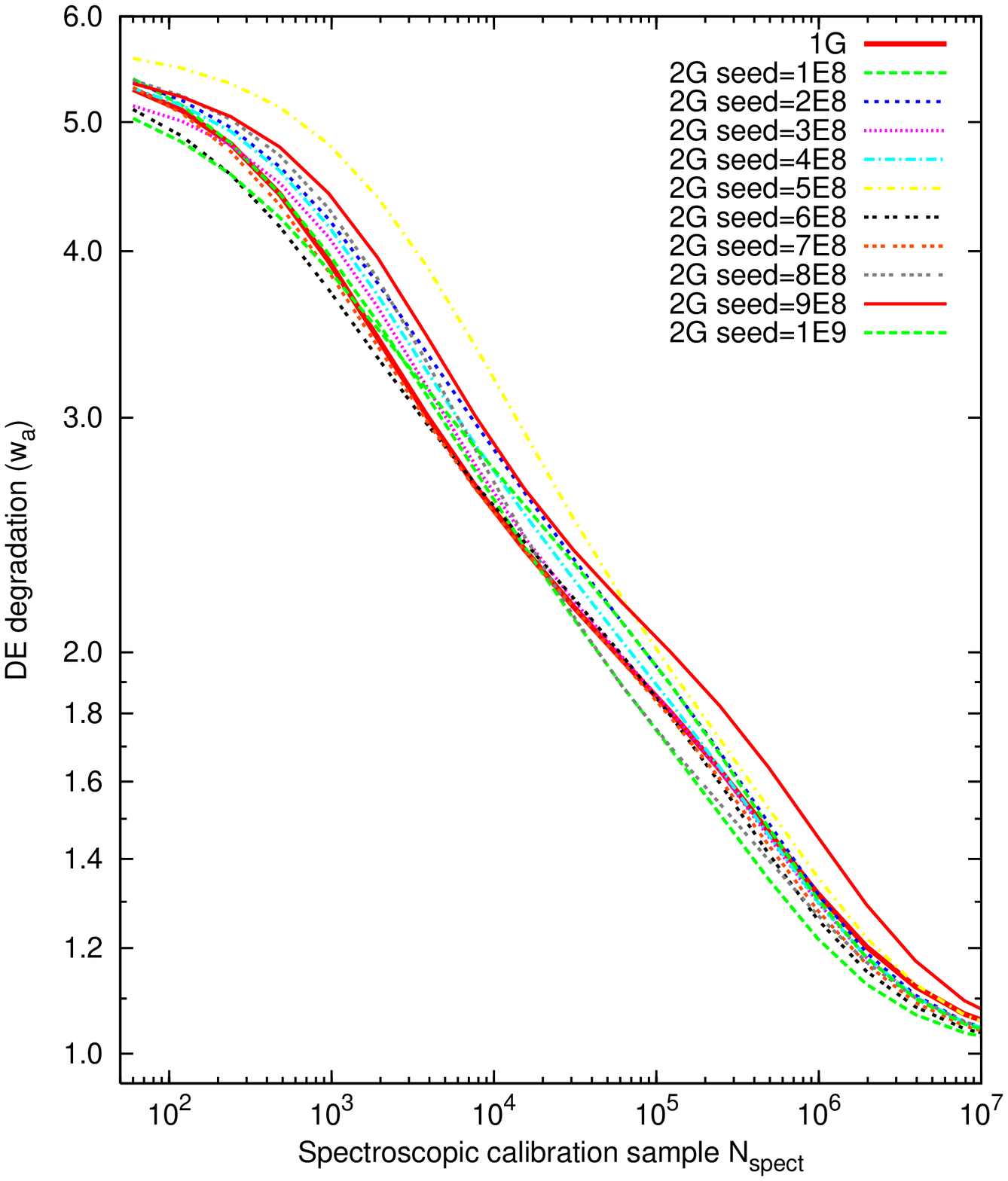, width=2.1in}
            \psfig{file=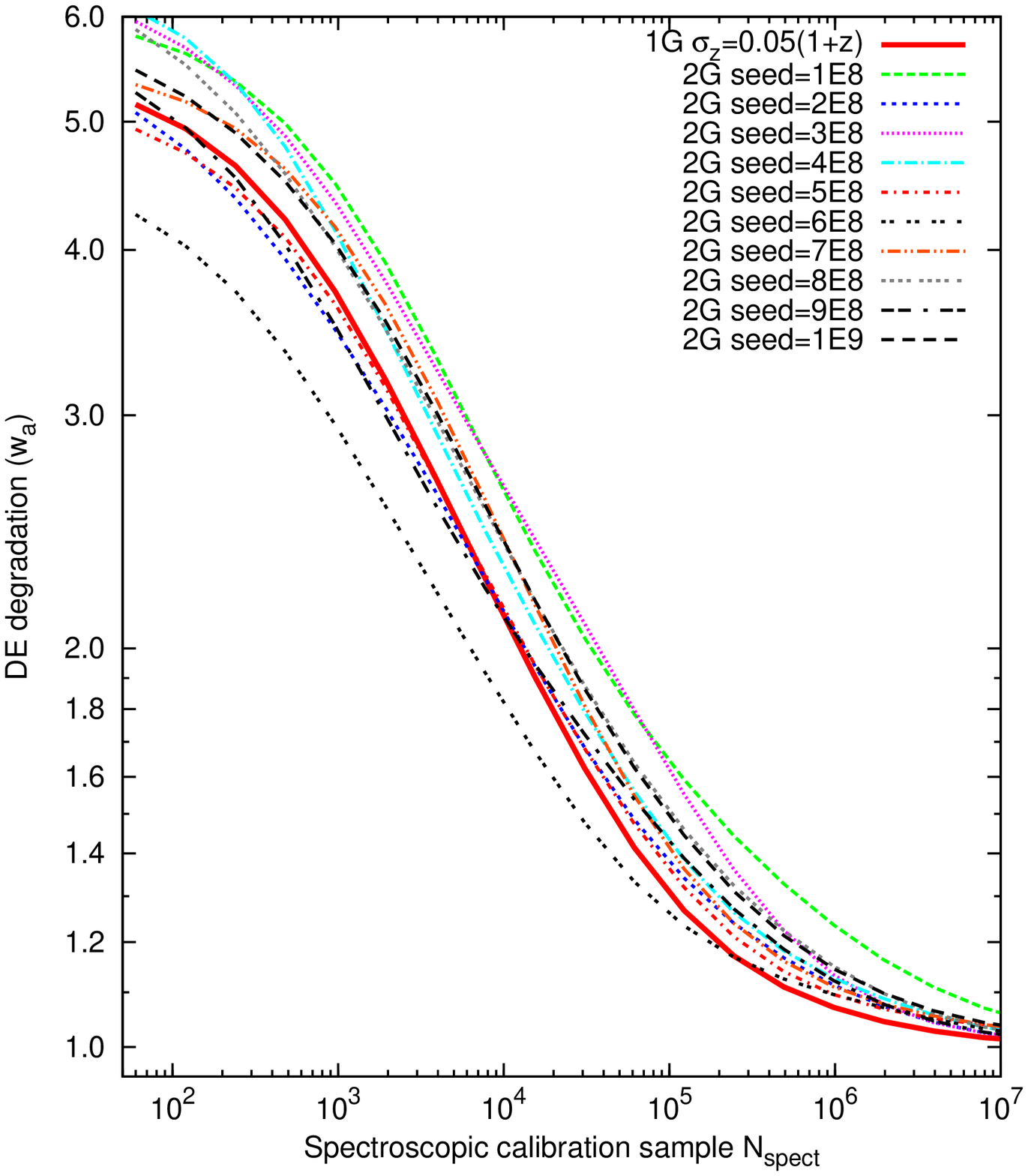, width=2.1in}}
\caption {The $N_{\rm spect}$ requirement for different fiducial photo-z
          models with $N_g = 2$, including some with randomly generated
          fiducial photo-z error distributions.
          Details of the photo-z models are in \S\,\ref{sec4.2}.
          Left: Information from the galaxy photo-z distribution
                      $n(z_{\rm ph})$ and the spectroscopic calibration
                      sample are used to constrain the underlying galaxy
                      redshift distribution $n(z)$ and photo-z parameters.
          Middle: Same as the left panel, except that information 
                      from $n(z_{\rm ph})$ is not utilized and $n(z)$ is 
                      assumed to be known {\it a priori}.
          Right: Within each of the three $\delta z = 1$ intervals,
                      the randomly generated fiducial $z_{\rm bias}$
                      and $\sigma_z$ values increase linearly with $1+z$.
                      The proportionalities are generated randomly.
          In all three panels, the thick solid red line is for the case of 
          a single Gaussian ($N_g = 1$).
          }
\label{fig:Nspect2G}
\end{figure*}

The left panel of Figure\,\ref{fig:Nspect2G} shows dark energy degradation
versus $N_{\rm spect}$ for several $N_g=2$ models,
all having fiducial rms width $0.05(1+z)$, but with different fiducial
biases and dispersions for the two components.
In detail, our study includes fiducial photo-z distributions in which:
the component Gaussians have
the same biases but different $\sigma_z$ values (``2G $\sigma_z$ diff'' model);
the same $\sigma_z$ values but different biases (``2G zbias diff'');
the same biases and $\sigma_z$ values but with
normalizations at a 3 to 1 ratio (``2G 3:1''); and 10 models in which
the fiducial $z_{\rm {bias;j}}$ and $\sigma_{z;j}$ are randomly
assigned while maintaining fixed rms photo-z error (``2G seed xxx'' models).

The $N_{\rm spect}$
requirements span a rather large range. For example, at $50\%$ dark energy
degradation, most of the photo-z models' $N_{\rm spect}$ requirement is
within a factor of 4 of that of the single-Gaussian model. But some
of the models require $40$ times more  $N_{\rm spect}$. 
Three- and four-Gaussian photo-z models exhibit similar behaviors.

    To understand the wide range of $N_{\rm spect}$ requirements for 
different photo-z models, we perform the following test.
We fix the underlying galaxy redshift ${n(z)}$ and do not use any 
information from ${n(z_{\rm ph})}$. 
The resulting $N_{\rm spect}$ requirements for the double-Gaussian 
photo-z models are shown in the middle panel of Figure\,\ref{fig:Nspect2G}.
At fixed dark energy degradation, the range of $N_{\rm spect}$ 
requirements is greatly reduced. For example, at $50\%$ dark
energy degradation, the $N_{\rm spect}$ requirement is within a factor of 2 
of that of the single-Gaussian model. We find
similar reduction of the range of $N_{\rm spect}$ requirements in the 
case of three- and four-Gaussian models.
   The test shows that the reason for the
wide range of $N_{\rm spect}$ requirements for different photo-z
models is that $n(z_{\rm ph})$ constrains the underlying galaxy 
redshift distribution and the photo-z parameters much better in 
some of the photo-z models than others. It is the redshift knowledge,
rather than weak-lensing information itself, 
that is sensitive to the details of the 
photo-z probability distribution.

   One possible cause of the poor sensitivity in some photo-z
models is the rapid variation of photo-z parameters in redshift.
The right panel of Figure\,\ref{fig:Nspect2G} shows the result of
reducing the degree of rapid variation of the photo-z parameters.
The range of $N_{\rm spect}$ is reduced to within a factor of
4 of that of the single-Gaussian model as shown in right panel
of Figure\,\ref{fig:Nspect2G}. 
In detail, we demand that the fiducial photo-z parameter
$z_{\rm bias}$ and $\sigma_z$ values to be proportional to  $1+z$
within each of the three redshift intervals with width $\delta z = 1$.
The proportionalities are generated randomly.
These photo-z models are much smoother than those randomly
generated in the left panel of Figure\,\ref{fig:Nspect2G}. This test
shows that $n(z_{\rm ph})$ is less effective in constraining the underlying
galaxy redshift distribution and photo-z parameters when the photo-z
model is rapidly varying.
In reality, photo-z parameters would most likely show smooth variations in 
redshift. The required calibration sample is expected to be within a factor
of a few times that of the single-Gaussian fiducial model.

We point out that multi-Gaussian cases
may require fewer spectroscopic calibration galaxies than the single-Gaussian
case. As an example, examine the photo-z model with double Gaussians
whose $\sigma_{\rm z}$ values are different. Its $N_{\rm spect}$ requirement
is shown in Figure\,\ref{fig:Nspect2G} (left) using the dotted blue line.
Since we keep the width
of $P(z_{\rm ph}|z)$ fixed, one of the Gaussians in the double-Gaussian 
photo-z model is narrower than the width of  $P(z_{\rm ph}|z)$ and
the other Gaussian is broader.
The narrower Gaussian tends to reduce the  $N_{\rm spect}$ requirement,
while the broader one tends to do exactly the opposite. The outcome of
these competing effects could
be either a smaller or larger requirement of the calibration sample. 
For this particular photo-z model, the required $N_{\rm spect}$ crosses 
that of the single-Gaussian model (shown as the thick solid red curve in
Fig\,\ref{fig:Nspect2G} left).

  We note that the generic behavior $\sigma_{w_a}\propto N_{\rm
  spect}^{0.2-0.25}$ continues to hold for all the fiducial
distributions, until the dark energy degradation drops to 1.2--1.3.  This
inflection typically occurs with a few times $10^5$ spectra, for the
LSST survey parameters assumed here.

\section{Optimizing the Spectroscopic Calibration Sample}
\label{sec:optimize}

So far we have been assuming that the calibration sample is 
uniformly distributed in
redshift. Weak lensing may require more precise photo-z calibration at some
redshifts than others. It could be beneficial if we distribute the 
calibration sample according to lensing sensitivity. 
Our goal is to find the $N_{\rm spect}^i$ that leads to the best dark energy
constraints for a fixed spectroscopic observing time $T_{\rm obs}$. This
could be modeled as
\begin{eqnarray}
  (\mathrm{Uncertainties} &&{} \mathrm{in\,\,dark\,\,energy\,\,parameters}) =
      \nonumber \\
  &&{} {\it function}(N_{\rm spect}^i \,,
      i = 1,2,...) \,,
\label{eqn:modelEqn}
\end{eqnarray}
\begin{equation}
 \sum_{i=1} N_{\rm spect}^i cost(z^i) = T_{\rm obs} \,,
\label{eqn:conEqn}
\end{equation}
where $cost(z^i)$ is the time it takes to obtain the spectrum of a galaxy at
redshift $z^i$. This is a constrained nonlinear optimization problem.
To calculate the function in equation\,\ref{eqn:modelEqn}, we first
calculate the Fisher matrices ${\rm F}^{\rm lens}$ and
${\rm F}^{n(z_{\rm ph})}$
for the presumed survey.  Then for each trial set of $N^i_{\rm
  spect}$, we calculate $F^{\rm spect}$ using
equation\,\ref{eqn:NspecPrior}, sum the Fisher matrices, and forecast
the dark energy uncertainties.
As to the constrain equation (\ref{eqn:conEqn}), we 
need to know the cost function. For illustrative purposes, we assume
$cost(z^i)$ is a constant.
\begin{figure}[ht]
\centerline{\psfig{file=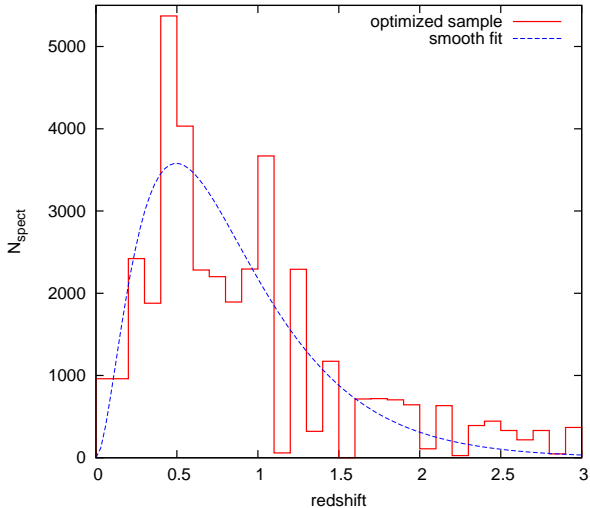, width=3.2in}}
\caption {Histogram: Optimal $N_{\rm spect}$ distribution in 
          redshift for single-Gaussian model. 
          Dark energy degradation is $56\%$ if this sample is distributed
          uniformly in redshift. The $N_{\rm spect}$ distribution in this 
          figure lowers dark energy degradation to $38\%$, which would
          require $69,000$ galaxy spectra to calibrate if the distribution
          is flat in redshift.
          Blue dashed line: Smooth fit to the histogram. The dark
          energy degradation is $44\%$ if this calibration sample is used.
          For both the histogram and the smooth fit, the calibration sample
          has 37,500 galaxies.
          }
\label{fig:optimalNspect}
\end{figure}

    As an example we choose a calibration sample of 37,500 galaxies and
assume a single-Gaussian photo-z model. If this calibration sample is 
uniformly distributed in redshift, dark energy degradation is $56\%$. 
If instead we use a downhill simplex method to find the
spectroscopic redshift sampling distribution that minimizes the dark
energy uncertainties for a fixed total number of redshifts, we obtain
the distribution shown as the 
histogram in Figure\,\ref{fig:optimalNspect}.  The optimized redshift
sampling lowers the dark energy
degradation to $38\%$. That is a $18\%$ gain in dark energy precision
at fixed investment of spectroscopy time.
From a different prospective, to reach $38\%$
dark energy degradation with a uniformly distributed calibration sample,
$69,000$ galaxy spectra are required. So optimization saves $46\%$
of the spectroscopic observing time for fixed cosmological degradation.
Multi-Gaussian photo-z models exhibit very similar behaviors.

We do not know exactly why the optimized calibration sample
distribution is not very smooth. It would be rather difficult to plan
the observation to match this distribution. Fortunately, a smooth
distribution like the one shown using the blue dashed line in 
Figure\,\ref{fig:optimalNspect} produces $44\%$ dark energy degradation, 
which is a moderate improvement over the uniform case.

\section{Conclusion and Discussion}
\label{sec:conclusion}

We explore the dependence of cosmological parameter uncertainties in
WL power-spectrum tomography on the size of the spectroscopic sample
for the calibration of photometric redshifts.  We present a formula that is
valid for arbitrary parameterizations of the photo-z error distribution and
then apply this to a multi-Gaussian model to see whether previous
works' assumptions of simple Gaussian photo-z errors were yielding
accurate results.

Indeed, we find that the required $N_{\rm spect}$ under the simple
Gaussian model is increased $\approx 4$ times when we allow more
freedom in the shape of the core of the photo-z distribution.
Fortunately, there appears to be an asymptotic upper limit as we add
more photo-z degrees of freedom.

   We also find a generic behavior $d\log\sigma / d\log N_{\rm
  spect}=$ 0.20--0.25, where $\sigma$ is the uncertainty in a dark energy
parameter, in the regime where $\sigma$ is degraded 1.2--5 times
compared to the case of perfect knowledge of the photo-z
distribution.  Hence, the fourfold increase in required $N_{\rm spect}$
from relaxing the Gaussian assumption is equivalent to a
$\approx 1.3$ times degradation in $\sigma$ at fixed $N_{\rm spect}$.

The exact value of dark energy degradation versus $N_{\rm spect}$ depends
significantly on the shape of the fiducial distribution, even when
the total rms photo-z error is held fixed.  For the case of the LSST
survey with rms photo-z error $0.05(1+z)$, we find that the ``knee''
at a dark energy degradation of 1.2--1.3 occurs in the range $N_{\rm
  spect}\approx10^5$--$10^6$.  

  For photo-z models described by nondegenerate Gaussians,
the size of the calibration sample varies by as much as 40 times among
the 14 models studied. Most of the variation is caused by the different
ability of the galaxy photo-z distribution $n(z_{\rm ph})$ to constrain
the underlying galaxy redshift distribution and the photo-z probability
distribution. These photo-z models whose parameters vary rapidly in
redshift are the ones that are least constrained. In reality, photo-z
parameters are expected to be smoothly varying in redshift.
The $N_{\rm spect}$ requirement would be only a factor
of a few from that of the single-Gaussian fiducial distribution.

    Finally, we show that the size of the calibration sample can be effectively
reduced by optimization. In a simple example, an optimized calibration sample
of 37,500 redshifts was able to reach the same dark energy degradation
as a sample of 69,000 galaxies uniformly distributed in redshift.

We restrict this study to the effect of the core of the photo-z distributions.
Catastrophic photo-z errors could potentially be very damaging. The methodology
provided in this study is applicable to study the effect of catastrophic
photo-z errors. We leave this to future work.

The methodology we use assumes that the spectroscopic survey is a
{\em fair} sample of the photo-z error distribution and is the {\em
  only} information available on the photo-z error distribution.
Since we have used a Fisher matrix technique, no photo-z estimation
method, regardless of technique (neural net, template fitting, etc.)
can surpass our forecasts under these conditions.

The calibration's success depends crucially on the spectroscopic
redshifts being drawn without bias from the redshift distribution of
the photometric sample it represents.
The survey strategy must be carefully formulated to make sure that
this occurs.  Differential incompleteness between, say, red and blue
galaxies or redshift ``deserts'', must be avoided.  This has not been
achieved by any large redshift survey beyond $z\approx 0.5$ to date.

It may be possible to constrain $P(z_{\rm ph}|z)$ by other means in
the absence of a fair spectroscopic sample of the size we specify.
One could invoke astrophysical assumptions, namely, that the spectra of
faint galaxies are identical to those of brighter galaxies, in an
attempt to bootstrap a fair bright sample into a calibration for
fainter galaxies.  Another suggestion
(\cite{Schneider06}; J. Newman, private communication) is
that the photometric sample be cross-correlated with an incomplete
spectroscopic sample to infer the redshift distribution of the
former.  It remains to be seen, however, whether these techniques can
attain the accuracy needed to supplant a direct fair sample of $>10^5$
spectra.  This would require some {\it a priori} bounds on the
evolution of galaxy spectra and the clustering correlation
coefficients of different classes of galaxies.  We look forward to
future progress in these techniques, keeping in mind that the demands
for precision cosmology from WL tomography are much more severe than
the demands that galaxy evolution studies typically place on
photometric redshift systems.

\acknowledgements {\it Acknowledgments}: We thank Wayne Hu, Dragon Huterer,
and Bhuvnesh Jain for useful discussions. Z.M. and G.B. are supported by
Department of Energy grant DOE-DE-FG02-95ER40893.
G.M.B. acknowledges additional support from NASA grant BEFS 04-0014-0018
and National Science Foundation grant AST 06-07667.

\renewcommand{\theequation}{A-\arabic{equation}}
\setcounter{equation}{0}
\section*{Appendix: Derivation of Equation\,14}
\label{sec:AppendixA}

   If one draws $N$ events from a sample with probability distribution function
$P(x;{\bft})$, where the components of ${\bft}$ are the parameters 
specifying the distribution and $x$ is the variable whose probability 
distribution is under consideration, what are the constraints on the
parameters ${\bft}$?

   Let us first divide $x$ into small bins and label the width of the bins as
$\Delta x_i$. The number of events that fall in the $i$th bin is Poisson
distributed with mean $\bar{N_i} = N P(x_i;{\bft}) \Delta x_i $.
The likelihood function can be expressed as
\begin{equation}
 L \propto \prod_i {\exp(-\bar{N_i}) \bar{N_i}^{N_i} \over {N_i !}} \,,
\end{equation}
and the natural logarithm of $L$ is,
\begin{equation}
 {\mathcal L} \equiv - ln L = \sum_i \bar{N_i} - N_i ln \bar{N_i} + ln N_i ! 
                              + const \,\,.
\end{equation}
The derivatives of ${\mathcal L}$ with respect to the model parameters
${\bft}$ are
\begin{equation}
 {\partial {\mathcal L} \over {\partial \theta_{\mu}}} = \sum_i
     \left ( 1- {N_i \over {\bar{N_i}}} \right )
     {\partial {\bar{N_i}} \over {\partial  \theta_{\mu}}} \,\,\,\,\,\,\,\,\, and,
\end{equation}
\begin{eqnarray}
 {\partial^2 {\mathcal L} \over {\partial \theta_{\mu} \partial \theta_{\nu}}} 
  &=&{} \sum_i \left [ {N_i \over {\bar{N_i}^2}} 
          {\partial {\bar{N_i}} \over {\partial  \theta_{\mu}}}
          {\partial {\bar{N_i}} \over {\partial  \theta_{\nu}}} \right.
   \nonumber \\
  &&{}
     \left.
     + \left ( 1- {N_i \over {\bar{N_i}}} \right )
     {\partial^2 {\bar{N_i}} \over {\partial  \theta_{\mu} 
      \partial \theta_{\nu}}} \right ] \,.
\end{eqnarray}
The Fisher matrix is,
\begin{eqnarray}
  F_{\mu \nu} &\equiv&{} \left <  
   {\partial^2 {\mathcal L} \over {\partial \theta_{\mu} \partial \theta_{\nu}}}
   \right > = \sum_i {1 \over {\bar{N_i}}}
          {\partial {\bar{N_i}} \over {\partial  \theta_{\mu}}}
          {\partial {\bar{N_i}} \over {\partial  \theta_{\nu}}}
        \nonumber \\
     &=& \sum_i {N \Delta x_i \over {P(x_i;{{\bft}})}}
          {\partial {P(x_i;{{\bft}})} \over {\partial  \theta_{\mu}}}
          {\partial {P(x_i;{{\bft}})} \over {\partial  \theta_{\nu}}}
        \nonumber \\
     &=& N \int dx {1 \over {P(x;{{\bft}})}}
          {\partial {P(x;{{\bft}})} \over {\partial  \theta_{\mu}}}
          {\partial {P(x;{{\bft}})} \over {\partial  \theta_{\nu}}} \,.
\label{eqn:Nprior}
\end{eqnarray}

   In the special case where $P(x;{{\bft}})$ is a Gaussian with mean
$\mu$ and spread $\sigma$,
\begin{equation}
  P(x;\mu,\sigma) = { 1 \over {\sqrt{2 \pi} \sigma} } \exp \left [
                    - {(x-\mu)^2 \over {2 \sigma^2}}
                    \right ] \,,
\end{equation}
we have
\begin{equation}
 {\partial P \over {\partial \mu}} = {x-\mu \over {\sigma^2}} P\,\,\,\,\,\,\,\,\,and,
\end{equation}
\begin{equation}
 {\partial P \over {\partial \sigma}} = -{P \over {\sigma}} +
                                  {(x-\mu)^2 \over {\sigma^3}} P \,.
\end{equation}
Plugging these results into equation\,\ref{eqn:Nprior} gives us
\begin{equation}
 F_{\mu \mu} = N \int_{-\infty}^{\infty} dx  {(x-\mu)^2 \over {\sigma^4}} P
             = {N \over {\sigma^2}} \,\,\,\,\,\,\,\,\, and,
\label{eqn:NpriorGmu}
\end{equation}
\begin{equation}
 F_{\sigma \sigma} = N \int_{-\infty}^{\infty} dx P \left [ 
               {(x-\mu)^2 \over {\sigma^3}} - {1 \over {\sigma}} \right ]^2
             = {2 N \over {\sigma^2}} \,.
\label{eqn:NpriorGsi}
\end{equation}
Note that $F_{\mu \sigma} = 0$ since the integral only involves odd powers of
$x-\mu$.

\end{document}